\documentclass[
	a4paper, 
	10pt, 
	unnumberedsections, 
	twoside, 
]{LTJournalArticle}
\usepackage{tabularx} 
\usepackage{amsmath}
\usepackage{multirow}%

\title{Perception-Guided EEG Analysis: A Deep Learning Approach Inspired by Level of Detail (LOD) Theory} 

\author{%
	BG Tong\textsuperscript{1}\thanks{Corresponding author: \href{mailto:dackmoon123@126.com}{dackmoon123@126.com}}
}

\date{\footnotesize\textsuperscript{\textbf{1}}Department of Psychiatry, Inner Mongolia People’s Hospital, Hohhot, China}

\renewcommand{\maketitlehookd} {%
\begin{abstract}
\textbf{Objective:} Analyzing the dynamic nature of human perceptual states from electroencephalogram (EEG) signals remains challenging for conventional deep learning methods. This study explores a novel approach, inspired by Level of Detail (LOD) theory, to address this by dynamically adjusting EEG data analysis to guide perceptual states. The core objective is to improve the identification accuracy of these states from EEG signals and to provide new avenues for personalized psychological therapy.
\textbf{Methods:} The research employs portable EEG devices to
collect data, combined with music rhythm signals for analysis. We
introduce the LOD theory to dynamically adjust the processing levels
of EEG signals, extracting core features related to perception. The
software system is developed using the Unity engine, integrating
audio materials and MIDI structures, and enabling the integration of
EEG data with Unity. The deep learning model includes a Convolutional
Neural Network (CNN) for feature extraction and classification, and a
Deep Q-Network (DQN) for reinforcement learning to optimize music
rhythm adjustment strategies.
\textbf{Results:} The CNN model achieved a 94.05\% accuracy in the
perceptual state classification task, demonstrating excellent
classification performance. The DQN model successfully guided
subjects' EEG signals to the target perceptual state with a 92.45\%
success rate on the validation set, requiring an average of 13.2
rhythm cycles to complete the state guidance. However, subjective
feedback from users indicated that approximately 50\% of the
researchers experienced psychological sensations corresponding to the
target state during the rhythm adjustment process, suggesting room
for improvement in the system's effectiveness.
\textbf{Discussion:} The experimental results validate the potential
of the LOD-based deep learning algorithm in EEG biofeedback and
perceptual guidance. Despite the preliminary achievements, the study
has limitations, such as the single source of the dataset, the
subjectivity of labels, and the need for further optimization of the
dynamic adjustment mechanism in the reinforcement learning reward
function. Future research will expand to diverse subject groups,
introduce more varied musical elements, and explore more advanced
reward functions to enhance the model's generalization ability and
personalized experience. \\
\textbf{Keywords:} EEG analysis, deep learning, Level of Detail (LOD)
theory, perceptual guidance, mental state classification
\end{abstract}
}

\begin{document}

\maketitle % Output the title section

%----------------------------------------------------------------------------------------
%	ARTICLE CONTENTS
%----------------------------------------------------------------------------------------

\section{Introduction}

\subsection{Research Background}
In the rapidly evolving fields of neuroscience and artificial
intelligence, the relationship between psychological phenomena and
neural signals remains an unsolved mystery. Although our
understanding of brain structure and neural activity has deepened,
inferring individual psychological experiences from microscopic
biological signals remains highly challenging. This involves both
technical difficulties and philosophical paradoxes, such as whether
psychological phenomena can be reduced to neural signals
\cite{luppi2021like}. This study proposes a new approach from electroencephalogram (EEG)
biofeedback to perceptual regulation, exploring the interaction
between music rhythm and neural activity to understand and guide
human psychological states.

In the field of EEG signal analysis and psychological phenomena
research, achievements and challenges coexist
\cite{bashivan2015learning}. Traditional studies often use machine
learning methods to identify EEG signal patterns to distinguish
mental states or diseases, such as epilepsy detection. These methods
extract time-domain and frequency-domain features and use traditional
classifiers for detection, but their effectiveness in classifying
complex psychological phenomena is limited
\cite{zhang2019deeplearningdecodingmental}. Although deep learning
methods have brought new opportunities and improved classification
accuracy in some studies, they still face limitations when dealing
with psychological phenomena, such as ignoring the complexity and
dynamics of perception and lacking consideration of individual
differences
\cite{roy2019deeplearningbasedelectroencephalographyanalysis}. In the
field of psychotherapy, EEG biofeedback technology has been explored,
but there are shortcomings in personalized treatment \cite{2019EEG}.

In contrast, this study aims to follow a different path. Drawing on
Gestalt psychology theory, it treats perception as the core unit of
psychological phenomena, breaking through traditional reductionist
thinking. By introducing Level of Detail (LOD) theory, the study proposes a new deep
learning method to dynamically adjust the processing levels of EEG
signals, improving the accuracy of perceptual state recognition and
providing new pathways for personalized psychological therapy. At the
same time, innovative data collection and processing methods are
adopted to advance the development of this field.

\subsection{Challenges in Modeling Psychological Phenomena}
Traditional deep learning methods, such as Convolutional Neural
Networks (CNNs) and Recurrent Neural Networks (RNNs), have shown
remarkable performance in pattern recognition and prediction tasks
\cite{bhatti2024comparativeanalysisdeeplearning}. However, they face
three main limitations when applied to psychological phenomena:
(1)Reductionism Issue: Psychological phenomena are diverse and
dynamic, making it difficult to represent them using a single data
stream or pattern; (2)Causality Assumption: Many models treat neural
signals as direct causes of psychological phenomena, ignoring the
possibility that knowledge may be a posteriori meaning structure;
(3) Individual Differences: The same EEG signal topology may produce
different subjective psychological experiences in different
individuals. These limitations serve as the motivation for the design
of this study, which will be elaborated in detail in the subsequent
sections \cite{10.3389/fncom.2024.1402689}.

\subsection{Inspiration and Reference from Gestalt Psychology}
\begin{figure} [h]
\centering
\includegraphics[width=0.9\columnwidth] {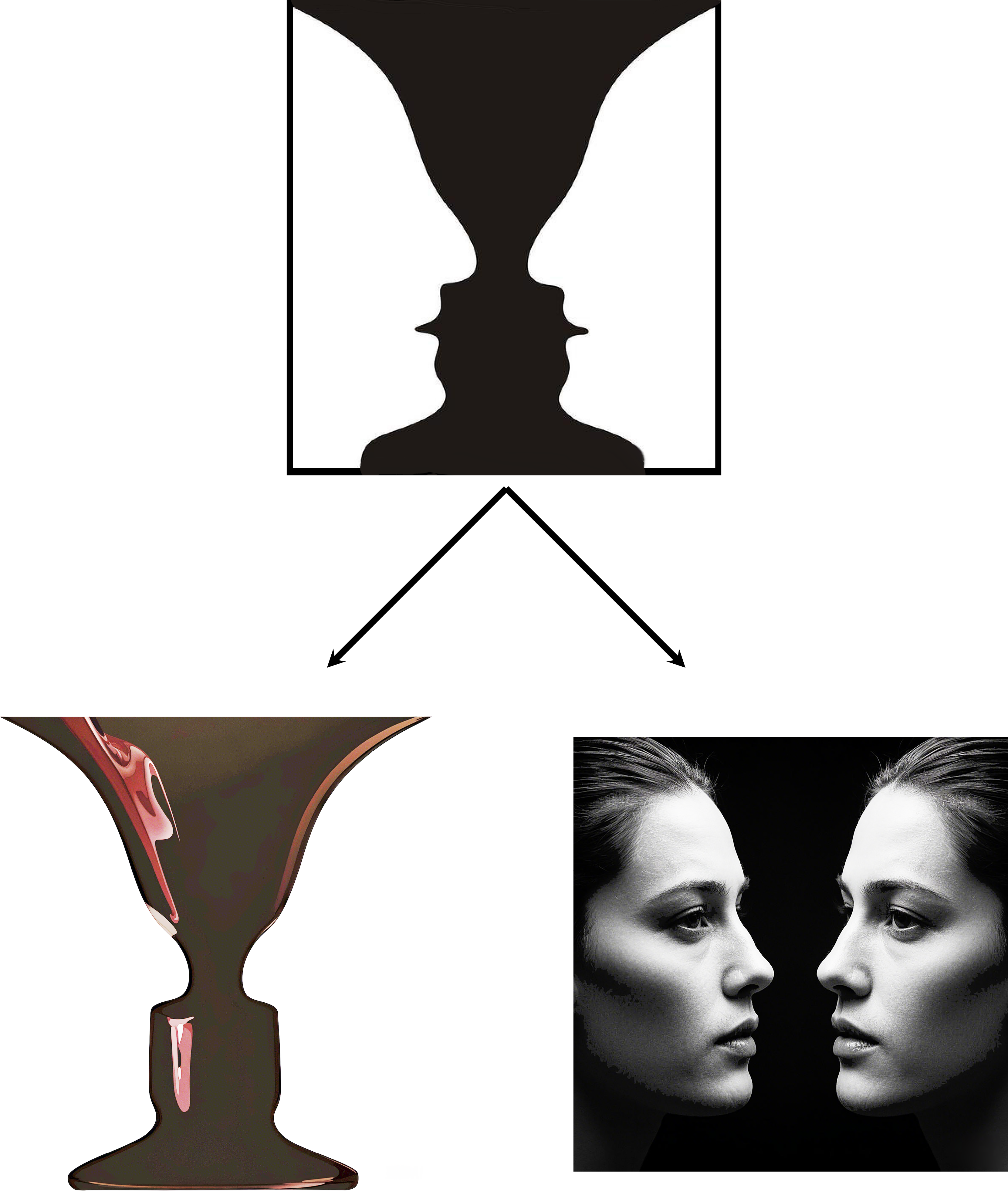}% 使用较小的比 % Note: Keep original image file name
\caption{Gestalt Perspective: Collapse of Brain Signal Structure into
Specific Perception.}\label{fig:Figure6} % <<< Changed label to fig:Figure1
\end{figure}

Gestalt psychology emphasizes the holistic nature of experience and
behavior, asserting that the whole is greater than the sum of its
parts. In visual perception, there are numerous organizational
principles, such as the laws of proximity and similarity. This holistic perspective is conceptually illustrated in Figure~\ref{fig:Figure6}. In the
field of deep learning for image recognition, this holistic
perspective aligns to some extent with how neural networks process
images by extracting features from the overall pixel distribution,
similar to how the human visual system organizes visual information
based on Gestalt principles
\cite{chen2024gestaltcomputationalmodelpersistent}. Additionally, the
concept of "insight" in Gestalt psychology shares a similar
underlying mechanism with the Transformer model's ability to predict
the next word in a sequence \cite{10.3389/fcomp.2024.1419831}.

\subsection{Recognition and Feedback of Biological Signals in Psychotherapy}
In the field of psychotherapy, the recognition and feedback of EEG signals hold significant importance. EEG signals contain rich psychological information, with different mental states exhibiting unique EEG patterns. By collecting EEG signals and converting them into feedback, combined with big data and artificial intelligence technologies, it is possible to construct psychological models, predict psychological changes, and intervene, thereby opening new avenues for psychotherapy. This study draws on Gestalt psychology theory, positioning perception as the fundamental unit of psychological phenomena and the integrator of information. A single-channel EEG device is used to collect data, aiming to extract meaningful perceptual features from complex background noise and develop training methods to optimize the understanding of overall perceptual states.

\subsection{Inspiration and Reference from LOD Theory}
The Level of Detail (LOD) concept was initially developed for 3D
modeling, where it optimizes resource allocation by dynamically
adjusting the level of detail in rendered objects
\cite{luebke2002level}. This study introduces LOD theory into
psychological modeling, treating EEG signals as multidimensional
dynamic data and capturing core features related to perception
through hierarchical processing
\cite{weng2024selfsupervisedlearningelectroencephalogramsystematic}. % <<< Added closing brace

In LOD, the camera's perspective determines the level of detail
displayed, analogous to an individual's subjective cognitive
framework. Different people may focus on different aspects of the
same neural signals, and the same person's cognitive framework may
vary across different moments or contexts. LOD reduces unnecessary
complexity by optimizing the processing of critical information,
similar to the "background-foreground separation" in perceptual
processes. The brain filters sensory information to form a holistic
perception, akin to the attention mechanism in deep learning. In deep
learning models, LOD can be applied to dynamic hierarchical
processing (low-level vs. high-level features), multi-perspective
generation (similar to "multi-camera perspectives"), and
simplification and generalization.

This study draws inspiration from LOD and applies it to the framework of deep learning algorithms, including the following operations:
\subsubsection{1.Importance-Based LOD:}
In techniques for optimizing 3D graphics rendering performance, importance-based LOD primarily considers the visual significance of objects in a scene. It not only focuses on the distance between the observer and the object but also takes into account factors such as the object's size, color, texture complexity, and position in the frame. In this study, greater computational resources are allocated to significant EEG features to enhance the model's sensitivity to key features. The data structure is designed as a two-dimensional combination of EEG data and audio data (with EEG data collected with a 200-millisecond delay). It is assumed that the important perceptual "points" depicted by LOD must coincide with the time points of audio-induced data, leveraging both the rhythm of audio nodes and the necessary "framing" of perception. The calculation formula is as follows:
     \begin{equation}  
          \begin{aligned} 
               & \hat{X} = X \cdot W_i 
          \end{aligned} 
          \label{eq:Formula-1}  
      \end{equation}
where $Wi$ represents the learned feature importance weights.
\subsubsection{2.Distance Threshold Method:}
The distance threshold method determines the level of detail based on the distance between the observer (camera) and the rendered object. When the distance between the object and the camera exceeds a specific threshold, the system switches to a lower level of detail. In this study, the distance threshold method is applied as a time-series operation to simplify input signals and simulate perceptual features at different time scales. When collecting EEG frequencies, an average of 256 data points is collected per loop. As the loop rate varies, the sampling interval changes accordingly, allowing the density of EEG feature data to dynamically adjust with the rhythm density. This better "outlines" the perceptual picture. The calculation formula is as follows:
     \begin{equation}  
          \begin{aligned} 
               & \hat{X} = \text{Pool}(X, k)
          \end{aligned} 
          \label{eq:Formula-2}  
      \end{equation}
where $X$ is the input data,$Pool$ represents the pooling operation, and $k$ is the pooling window size.
\subsubsection{3.Dynamic LOD:}
Compared to methods using predefined distance thresholds, dynamic LOD technology allows for smoother adjustments to the level of detail of objects. It can continuously adjust the complexity of objects within a certain range, including vertex reduction and texture simplification. Dynamic LOD automatically adjusts the level of detail based on real-time conditions (such as changes in perspective, hardware performance, and scene complexity) to maintain optimal performance and visual quality. This study posits that perceptual experiences exhibit similar high-dimensional spatial dynamics in human consciousness. Therefore, reinforcement learning is introduced into the algorithm design. It is observed that while subjects believe they are immersed in a rhythm and maintain a certain perception during EEG data collection, their internal perspective actually undergoes changes. For example, the current perceptual experience is a perception of previous perceptions, and the EEG-depicted image resembles a "picture within a picture," echoing the level of detail and focal points influenced by object distance in dynamic LOD. By leveraging relaxation and focus data provided by the TGAM chip, deep learning is used for correlation analysis (score-based reinforcement unsupervised learning). The calculation formula is as follows:
     \begin{equation}  
          \begin{aligned} 
               & \hat{X} = f(X) \cdot \sigma(W_e \cdot E)
          \end{aligned} 
          \label{eq:Formula-3}  
      \end{equation}
where $E$ represents relaxation and focus levels, $W_e$ is the weight matrix,and $\sigma$ is the activation function.
\subsubsection{4.Screen Space Error (SSE):}
Screen Space Error is used to evaluate the geometric difference between a simplified 3D model and the original high-detail model, ensuring that the simplified model's screen projection remains visually consistent with the original model and avoids noticeable discrepancies. This study hypothesizes that perceptual phenomena are essentially the reprojection of signal topologies generated by brain neurobiochemical processes (to be discussed in detail later). In the specific algorithm implementation, a CNN module is introduced, treating the subject's perceptual experience during rhythm loops as "fragile fragments." The collected EEG feature values are transformed from time series into a 256×4×2 matrix and input into an adjusted CNN convolutional neural network. The calculation formula is as follows:
     \begin{equation}  
          \begin{aligned} 
               & \text{SSE} = \frac{1}{N} \sum_{i=1}^{N} (X_i - \hat{X}_i)^2
          \end{aligned} 
          \label{eq:Formula-4}  
      \end{equation}

% ===== Add this paragraph for paper structure outline HERE =====
The remainder of this paper is organized as follows. Section Methods~\ref{sec:Methods} details the experimental methodology, including data collection, preprocessing, software design, label assignment, and the proposed deep learning model architecture incorporating LOD principles, as well as the reinforcement learning framework for perceptual guidance. Section Results~\ref{sec:Results} presents the experimental results for both perceptual state classification and guidance tasks. Finally, Section Discussion~\ref{sec:Discussion} discusses the findings, experimental limitations, philosophical reflections inspired by the study, and future research directions.

\section{Methods}\label{sec:Methods}

\subsection{Data Collection and Preprocessing}
\subsubsection{Data Collection}
This study adopts an exploratory approach, focusing on collecting EEG data from a single researcher % <<< MODIFIED: Changed to single researcher
acting as the subject. The aim is to explore individualized psychological perception
and EEG characteristics from the perspective of Gestalt psychology, acknowledging the limitations % <<< ADDED: Acknowledging limitations
of single-subject analysis regarding generalizability. Given the uniqueness of individual psychological experiences,
a portable EEG device equipped with a ThinkGear ASIC Module (TGAM) chip % <<< MODIFIED: Added explanation for TGAM
was used for data collection to capture these unique EEG patterns and their underlying psychological structures, specifically from the FP1 electrode position. % <<< ADDED: Specified FP1 channel

The experiments were conducted in a quiet daily environment, with the
researcher wearing the TGAM chip device for data collection. The TGAM chip % Use "TGAM chip" or "TGAM device" consistently
plays a crucial role in this study's data collection process. It not
only records raw data in real time but also provides spectral
information for $\alpha$, $\beta$, $\theta$, and $\delta$ waves, as well as two key indicators provided by the chip: Attention and Meditation. These indicators range from 0 to 100, with higher values indicating greater relaxation
or focus, providing an important basis for quantitatively assessing
the psychological state of the subject, see Table % <<< MODIFIED: changed "subjects" to "subject"
\ref{tab:byte_value}.

\begin{table}[t]
    \centering
    \resizebox{\columnwidth}{!}{%
    \begin{tabular}{ccl}
    \toprule
    \textbf{Byte} & \textbf{Value} & \textbf{Explanation} \\
    \midrule
    %0 & 0xAA & [SYNC] \\
    1 & 0xAA & [SYNC] \\
    2 & 0x20 & [PLENGTH] (payload length) of 32 bytes \\
    3 & 0x00 & [POOR\_SIGNAL] Quality \\
    4 & 0x83 & No poor signal detected (0/200) \\
    8 & 0x18 & [ASIC\_EEG\_POWER\_INT] \\
    9 & 0x20 & [VLENGTH] 24 bytes \\
    10 & 0x00 & (1/3) Begin Delta bytes \\
    11 & 0x00 & (2/3) \\
    12 & 0x94 & (3/3) End Delta bytes \\
    13 & 0x00 & (1/3) Begin Theta bytes \\
    14 & 0x00 & (2/3) \\
    15 & 0x42 & (3/3) End Theta bytes \\
    16 & 0x00 & (1/3) Begin Low-alpha bytes \\
    17 & 0x00 & (2/3) \\
    18 & 0x0B & (3/3) End Low-alpha bytes \\
    19 & 0x00 & (1/3) Begin High-alpha bytes \\
    20 & 0x00 & (2/3) \\
    21 & 0x64 & (3/3) End High-alpha bytes \\
    22 & 0x00 & (1/3) Begin Low-beta bytes \\
    23 & 0x00 & (2/3) \\
    24 & 0x4D & (3/3) End Low-beta bytes \\
    25 & 0x00 & (1/3) Begin High-beta bytes \\
    26 & 0x00 & (2/3) \\
    27 & 0x3D & (3/3) End High-beta bytes \\
    28 & 0x00 & (1/3) Begin Low-gamma bytes \\
    29 & 0x00 & (2/3) \\
    30 & 0x07 & (3/3) End Low-gamma bytes \\
    31 & 0x00 & (1/3) Begin Mid-gamma bytes \\
    32 & 0x00 & (2/3) \\
    33 & 0x05 & (3/3) End Mid-gamma bytes \\
    34 & 0x04 & [ATTENTION] eSense \\
    35 & 0x0D & eSense Attention level of 13 \\
    36 & 0x05 & [MEDITATION] eSense \\
    37 & 0x3D & eSense Meditation level of 61 \\
    38 & 0x34 & [CHKsUM] (1’s comp inverse of 8-bit \\ 
       &      & Payload sum of 0xCB) \\
    \bottomrule
    \end{tabular}%
    }
    \caption{The data structure in EEG packets sent by TGAM Bluetooth}
    \label{tab:byte_value}
    \end{table}

Each data collection session lasted 6 to 8 minutes and included 200
rhythm loops. The rhythm speed could be adjusted between 90 and 140
BPM, with each adjustment made in increments of 5 BPM. For example,
when the speed was set to 120 BPM, each rhythm loop lasted 2 seconds
(60 seconds / 120 BPM = 0.5 seconds per beat, with 4 beats forming a
loop, thus each loop lasting 2 seconds) \cite{Yedukondalu2025}.
Before each session, the researcher selected a seasonal label
(spring, summer, autumn, winter) reflecting their current % "their" should refer to the single researcher's state
psychological state, which served as the training label for the deep
learning model. A total of 418 independent data collection sessions
were completed for this subject, accumulating a rich dataset of paired EEG and music % <<< MODIFIED: Added "for this subject"
rhythm data.
\subsubsection{Data Preprocessing}
First, the dataset, comprising data from the single subject, was divided into training and test sets using an
80\%-20\% split. If necessary, a validation set could also be
created. This division helps evaluate model performance and prevent
overfitting. Each sample contains the following information (where N denotes the total number of samples): % <<< Added clarification for N

\begin{itemize}
\item EEG Data: Shape (N, 256, 4, 2), representing the raw time-series EEG signal structures for each loop. % Clarified "raw time-series"
% \item Focus and Relaxation Scores: Shape (N, 2), recording the focus and relaxation scores for each sample. % This structure might differ depending on how you use it later. Let's keep as is for now.
\item Attention and Meditation Scores: Shape (N, 2), recording the Attention and Meditation Scores scores provided by the TGAM chip for each sample. % Kept original wording
\item Drum Parameters: Shape (N, 16), including drum speed and drum
arrangement codes.
\item Perceptual Labels: Shape (N, 2), recording the state and its
percentage (spring, summer, autumn, winter).
\item Operation Records: Shape (N, 1), recording adjustments to speed
and density.
\end{itemize}

EEG data preprocessing is a critical step before model training. To
preserve the maximum amount of temporal information, raw
EEG data was collected without frequency-domain transformation. The sampling frequency
was set to 512 Hz, with 256 data points collected per rhythm loop,
ensuring precise alignment with the temporal structure of the music
rhythm.

However, since the TGAM dry electrode chip was used, EEG signals could be % Changed "may be" to "could be" for past tense consistency
affected by noise and artifacts, such as eye movement artifacts
(blinking) and power line interference (50/60 Hz). To ensure signal
quality and improve model accuracy, specific preprocessing steps were employed:
\begin{itemize}
    \item Low-pass filtering: A low-pass filter with a cutoff frequency of [Specify Cutoff Frequency, e.g., 45 Hz] % <<< SPECIFY CUTOFF FREQUENCY HERE
    was applied to remove high-frequency noise while preserving relevant EEG bands.
    \item Artifact Removal: Independent Component Analysis (ICA) % <<< Defined ICA here
    techniques were employed to identify and remove components associated with eye blinks and other artifacts, enhancing signal purity.
\end{itemize}

For audio processing, music rhythm signals were synchronously
collected, and the average audio value within the first 200
milliseconds of each rhythm loop was calculated as a feature input.
This approach effectively captures the influence of rhythm on EEG
signals while providing the model with contextual audio information.

Finally, EEG and audio data were integrated into a 256×4×2 data
structure. This structure cleverly incorporates time-series
information and corresponding audio features, serving as the input
for the subsequent Convolutional Neural Network (CNN) model. This
lays a solid foundation for the model to accurately analyze the
relationship between EEG signals and music rhythm
\cite{pandey2021brain2depthlightweightcnnmodel}.

\subsection{Software Design and Development}
The software system for this study was developed using the Unity
engine, aiming to achieve interactive EEG biofeedback and music
rhythm modulation. This design plays a crucial connecting role in the
overall research methodology, tightly integrating data collection,
model training, and user experience.

The software system incorporates several key elements. For audio
materials and Musical Instrument Digital Interface (MIDI) structure, MIDI drum materials generated by % <<< Defined MIDI
software such as BandInBox 2015, Addictive Drums 2, Studio Drummer,
and EZdrummer 2 were used, totaling approximately 4GB. These
materials cover a variety of music styles, including jazz, rock, and
hip-hop, and exhibit a wide range of drum complexity, from simple to
intricate. Each MIDI file is pre-labeled with important information
such as Beats Per Minute (BPM), drum density, and style tags. % <<< Defined BPM (assuming not defined before)
This enables the system to dynamically adjust music rhythms based on user
feedback, providing a rich foundation for personalized music
modulation based on user EEG data and psychological states.

\begin{figure*}
    \centering
    \includegraphics[width=1\textwidth]{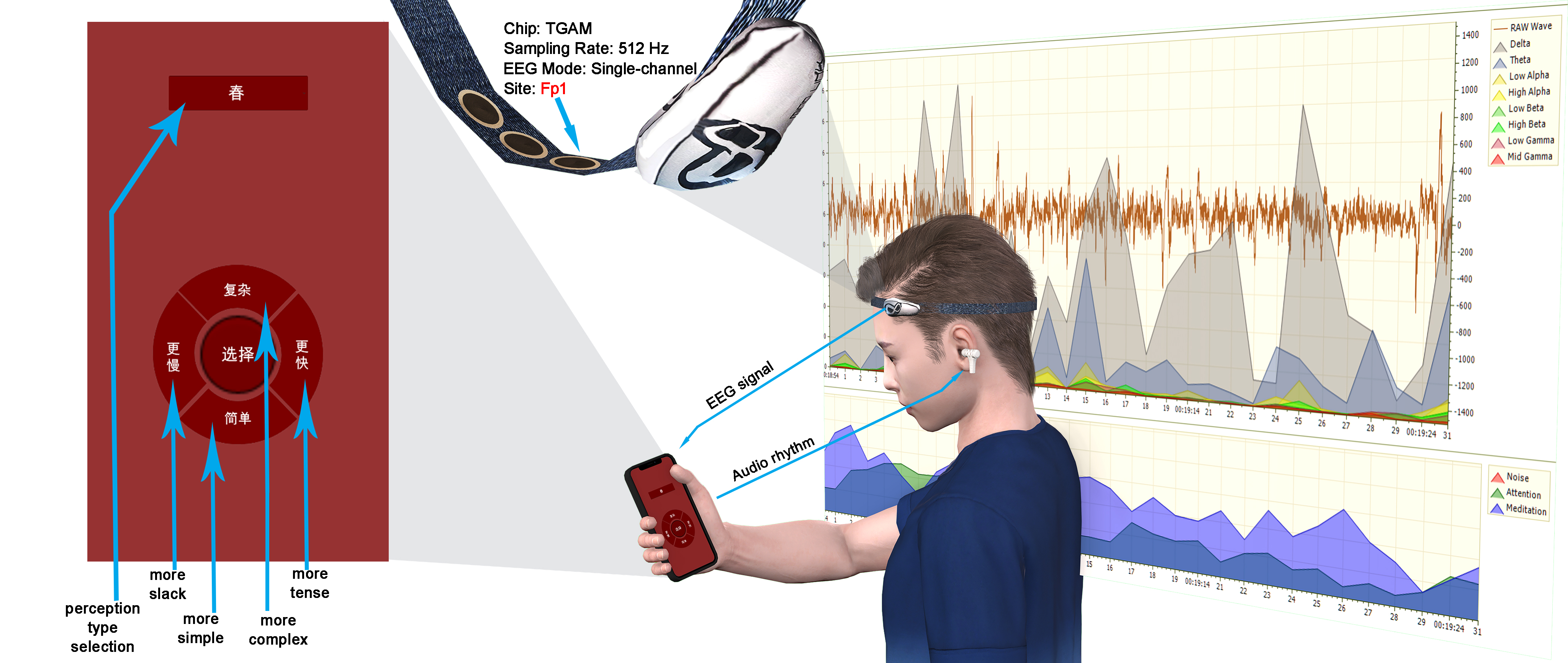}
    \caption{Schematic Diagram of the EEG Data Acquisition and Music Rhythm Modulation System}
    \label{fig:rhythm modulation}
\end{figure*}

The user interface and interaction design were implemented using
Unity's Graphical User Interface (GUI) system, featuring two adjustment axes: tension-relaxation % <<< Defined GUI
and complexity-simplicity. Users can click buttons on the interface,
such as "more tense," "more relaxed," "more complex," or "more
simple," to adjust drum rhythms in real time. After each adjustment,
the system automatically selects suitable drum combinations from the
MIDI library and plays them using Unity's audio engine, see Figure
\ref{fig:rhythm modulation}. This real-time interactive design not
only enhances users' ability to autonomously modulate music rhythms
but, more importantly, creates diverse experimental conditions for
collecting EEG data related to different psychological states through
users' active operations. This allows the research to more
comprehensively explore the complex relationship between music rhythm
and psychological states.

For the integration of EEG data with Unity, the TGAM chip establishes
a stable connection with an Android device running the Unity program
via Bluetooth, enabling real-time transmission of EEG data. Upon
receiving the EEG data, the Unity program immediately synchronizes it
with the current drum rhythm, generating a specific 256×4×2 data
structure. This data structure serves as the input for the deep
learning model, ensuring that the model can simultaneously analyze
the relationship between EEG signals and music rhythms. This provides
robust data support for accurately identifying and guiding perceptual
states.

Additionally, Unity's built-in reinforcement learning (RL) module (specifically, the ML-Agents toolkit) % <<< Defined RL and explained ML-Agents
was ingeniously applied to optimize drum rhythm adjustment
strategies. The model dynamically adjusts drum parameters based on
user EEG data (including key indicators such as Attention and Meditation scores)
and subjective feedback (e.g., clicking the "adapt to current state"
button). Through continuous learning and optimization, the system
gradually guides users toward the target perceptual state, achieving
intelligent interaction between the software system and users. This
provides an innovative data-driven approach to music rhythm
modulation for the research.

\subsection{Data Collection and Label Assignment}\label{sec:LabelAssignment} % Optional label for section
To ensure the validity of the data and the high quality of model
training, we carefully designed a dynamic label assignment mechanism
that runs through the entire data collection process. This mechanism
is of great significance for accurately reflecting changes in the user's % Changed "users'" to "the user's" for single subject
perceptual states.

\subsubsection{Initial State Selection and Dynamic Labeling}\label{sec:InitialState} % Changed title for clarity
At the beginning of each experiment, the subject (the researcher) was required to select a seasonal label % Clarified "subject"
("spring," "summer," "autumn," or "winter") that reflected their
current psychological state as the initial perceptual state. Data
collection then continued for 200 rhythm loops (approximately 6-8
minutes). During the data collection process, the drum rhythm speed
could be flexibly adjusted between 90 and 140 BPM, with each adjustment
made in increments of 5 BPM. This ensured that EEG data was collected
under different rhythm conditions. For example, when the drum rhythm
speed was set to 120 BPM, each rhythm loop lasted 2 seconds. This
standardized collection process facilitated subsequent data
processing and analysis.

% --- Content from the removed subsubsection title moved here ---
The core part of this mechanism is the state transition and label assignment during the session. If the
subject remained in the initially selected perceptual state at the end
of the experiment, all collected data would be explicitly labeled as % Changed "will be" to "would be" for hypothetical consistency
100\% initial state. For example, if the initial state was "autumn"
and no state change occurred by the end of the experiment, all data
would be labeled as 100\% autumn. However, if the subject transitioned
from the initial state (e.g., "autumn") to another state (e.g.,
"summer") during the experiment, the data labels would gradually
change from 100\% initial state to 100\% target state, forming a
linear gradient process \cite{Craik_2019}. This dynamic label
assignment method precisely records the trajectory of the subject's
perceptual state changes during the experiment, providing the
subsequent deep learning model with rich, time-series-labeled
information. This helps the model better learn and understand the
patterns of perceptual state transitions \cite{ding2024eeg}.

\subsubsection{Data Storage and Usage}Each experiment generates 50 data structures of size 256×4×2. Each data structure contains important information such as EEG features, audio features, Attention and Meditation scores, and perceptual state labels. These data comprehensively record various physiological and psychological indicators of the subject during the experiment and accurately reflect changes in perceptual states through the labels. These rich data are used to train and validate the deep learning model, ensuring that the model can accurately identify and guide perceptual states. This provides a solid data foundation for the entire study.

\subsection{Deep Learning Model Design}
The deep learning model designed in this study aims to extract features from EEG signals and music rhythms for the classification of perceptual states. The overall structure of the model consists of three main parts: data input and preprocessing, feature extraction and fusion, and perceptual state classification \cite{10.3389/fnhum.2024.1442398}.

\subsubsection{Data Input and Preprocessing}
\begin{itemize}
\item The model input is a tensor with dimensions 256×4×2 for each perceptual segment processed, representing the EEG and audio signals associated with a cycle of rhythm loops. The dimensions signify: % Simplified opening
    \begin{itemize}
    \item Time Dimension: 256 data points per loop, ensuring alignment with the temporal structure of the music rhythm.
    \item Loop Dimension: 4 consecutive rhythm loops forming a complete cycle, capturing periodic features.
    \item Feature Dimension: 2 channels integrating the relevant EEG signal information and the corresponding audio features (e.g., rhythm characteristics like BPM and density) for that time point and loop. The exact composition of these integrated features is determined during the feature engineering process based on the raw data and derived metrics (like band power). 
    \end{itemize}

\item Data Preprocessing Steps performed before model training:
    \begin{itemize}
    \item Normalization: Normalize EEG and audio features to unify data scales, preventing features with larger values from dominating the learning process. % Added rationale
    \item Data Augmentation: Introduce time-axis shifting and noise perturbation to the training data. This enhances model robustness and generalizability by exposing it to a wider variety of signal variations. % Added rationale
    \end{itemize}
\end{itemize}

\subsubsection{Feature Extraction and Fusion Based on LOD Theory}
\begin{itemize}
\item CNN Architecture: A multi-layer CNN is used to extract
spatiotemporal features. The structure involves: % Simplified structure description
    \begin{itemize}
    \item First Layer: 1×1 convolution to capture local features, reduce
    parameters, and output 32 channels, followed by Rectified Linear Unit (ReLU) activation. % <<< Defined ReLU
    \item Second Layer: 3×3 convolution followed by max pooling to extract spatial features of
    time and audio signals. This is followed by ReLU activation and 2×2 max
    pooling to further extract high-level features, capturing the complex relationship between EEG signals and
    music rhythms. % Simplified description assuming only one 3x3 layer block
    \end{itemize}
\item LOD Fusion:
    \begin{itemize}
    \item Feature Fusion: After feature extraction by the convolutional
    layers, the resulting EEG and audio features are concatenated along the channel
    dimension to form a unified feature representation before applying LOD mechanisms. % Added timing clarification
    \item LOD Application: Different LOD concepts are applied to modulate the feature representation:
        \begin{itemize}
        \item Importance-Based LOD: Assign weights ($W_i$) based on learned feature
        importance to enhance the model's sensitivity to key perceptual features. These weights are typically learned during the model training process, often guided by the task objective (e.g., classification accuracy) or specific attention mechanisms within the network. The operation is defined by $\hat{X} = X \cdot W_i$ \eqref{eq:Formula-1}, where $X$ represents the input features to this stage and $\hat{X}$ represents the weighted features. % <<< Added explanation for Wi, defined X, X_hat
        \item Distance Threshold Method (Simulated): Simplify signals through time-series
        pooling operations (e.g., max pooling or average pooling) to simulate perceptual features at different time
        scales, analogous to reducing detail based on distance. The formula is $\hat{X} = \text{Pool}(X, k)$ \eqref{eq:Formula-2}, where $X$ is the input data, $Pool$ represents the pooling operation, and $k$ is the pooling window size. % <<< Pool defined here is sufficient as X and X_hat defined above
        \item Dynamic LOD: Introduce reinforcement learning or attention mechanisms to dynamically
        adjust feature detail levels based on real-time conditions, such as predicted task relevance or auxiliary inputs like Attention/Meditation scores ($E$). The formula is $\hat{X} = f(X) \cdot \sigma(W_e \cdot E)$ \eqref{eq:Formula-3}, where $f(X)$ represents the features extracted by the preceding layers (e.g., the CNN), $E$ represents auxiliary context like Attention/Meditation levels, $W_e$ is a learned weight matrix for the context, and $\sigma$ is an activation function. % <<< Defined f(X)
        \end{itemize}
    \end{itemize}
\end{itemize}

\subsubsection{Perceptual State Classification}
\begin{itemize}
\item Fully Connected Layer: Map the feature vectors extracted by the
CNN and potentially modulated by LOD mechanisms % Added context link to previous step
to perceptual state classifications (spring, summer, autumn,
winter). ReLU activation is used to introduce non-linearity.
\item Output Layer: Use the softmax function to generate probability
distributions over the perceptual states. The softmax function ensures that the outputs are non-negative and sum to one, representing the model's confidence for each of the four possible states. The final output is the probability of each state. % <<< Added optional explanation for softmax
\end{itemize}

\subsubsection{Model Training and Optimization}
The model training process adopts a supervised learning strategy,
with the following specific steps:
\begin{itemize}
\item Loss Function: Use the cross-entropy loss function to optimize
the classification task. The loss function is defined as:
    \begin{equation}
    \begin{aligned}
    & L = -\frac{1}{N} \sum_{i=1}^{N} \sum_{j=1}^{C}y_{ij} \log(p_{ij})
    \end{aligned}
    \label{eq:Formula-8}
    \end{equation}
    where $L$ is the loss function, $N$ is the number of samples, $C$ is
    the number of categories (4 in this study, corresponding to the four
    seasons), $y_{ij}$ is a binary indicator (0 or 1) indicating if category $j$ is the correct classification for sample $i$, % <<< CORRECTED explanation for y_ij
    and $p_{ij}$ is the model's predicted probability of the $i$-th
    sample belonging to the $j$-th category.
\item Optimizer:
    \begin{itemize}
    \item Use the Adam optimizer for parameter updates, with an initial
    learning rate of 0.001.
    \item The Adam optimizer adaptively adjusts the learning rate, making
    it suitable for handling non-stationary data distributions.
    \end{itemize}
\item Training Strategy:
    \begin{itemize}
    \item Divide the dataset into training and test sets using an 80\%-
    20\% split to prevent overfitting. A separate validation set (10-20% of the training data) was used for hyperparameter tuning and early stopping. % Added mention of validation set use
    \item The model was trained for a maximum of 200 epochs with a batch size of 32. % <<< FILLED with provided values
    \item Use early stopping to halt training when the validation loss no
    longer decreased for a predefined number of epochs (10 patience epochs), selecting the model weights that achieved the best validation performance. % Added detail on early stopping mechanism
    \end{itemize}
\end{itemize}

\subsection{Reinforcement Learning Framework}
\begin{figure}[h]
    \centering
    \includegraphics[width=0.9\columnwidth]{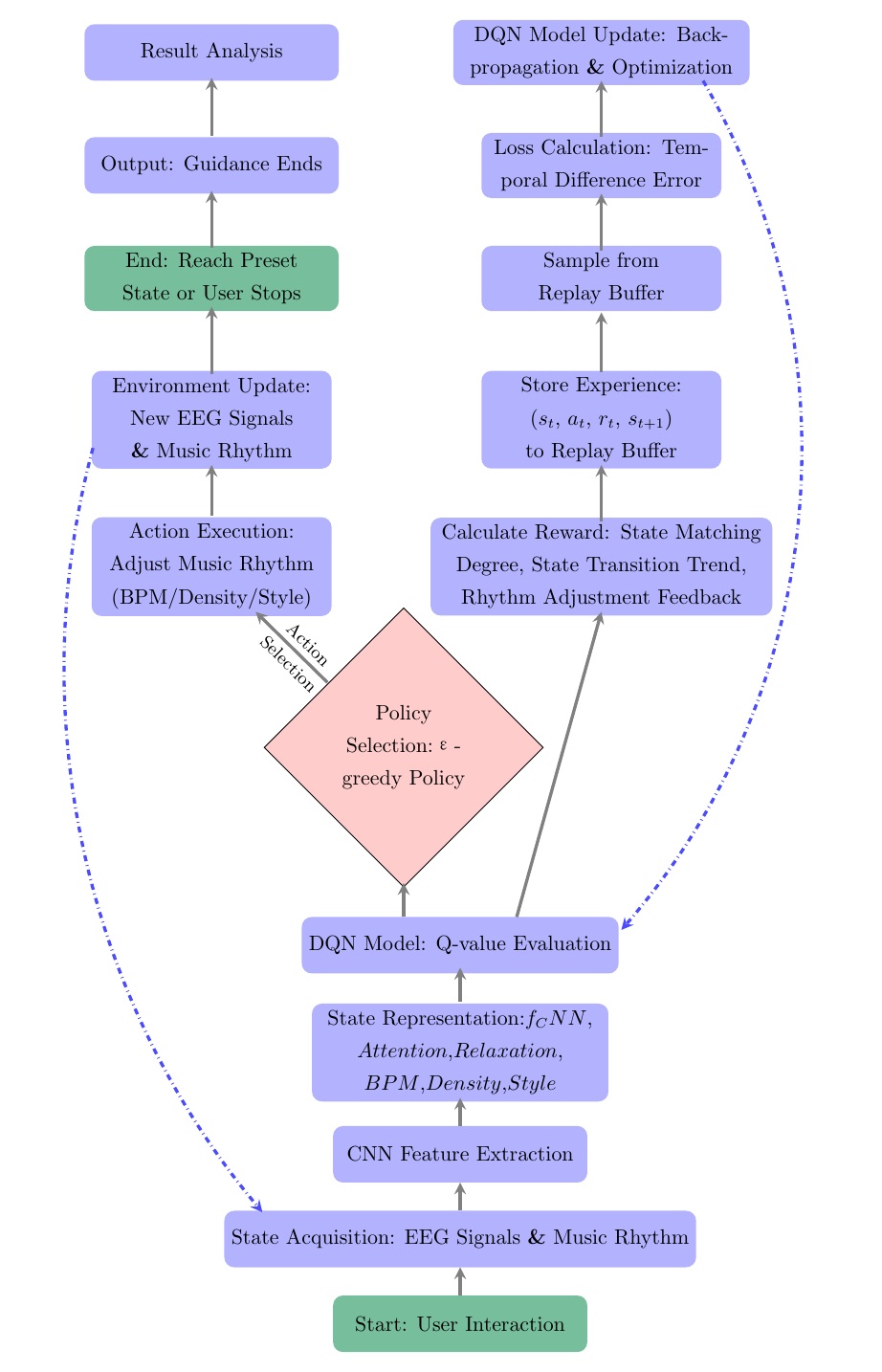} % 使用较小的比例适应一列宽度
    \caption{Flowchart of the DQN-based Perceptual State Guidance System.}\label{fig:CNN+Q}
\end{figure}

The reinforcement learning module dynamically adjusts music rhythms
to guide the user from their current perceptual state to the target % <<< Changed "users" to "the user"
state. This study employs a Deep Q-Network (DQN) combined with
features extracted by the CNN and user feedback to optimize rhythm
adjustment strategies \cite{10220592}, see Figure \ref{fig:CNN+Q}.

\subsubsection{State Space Definition}
The state space represents the current environmental information of
the system, including:
\begin{itemize}
\item CNN-Extracted Features: Spatiotemporal features extracted from
the 256×4×2 input data, representing the complex relationship
between EEG signals and music rhythms.
% --- MODIFIED Item Description ---
\item Attention and Meditation Scores: Attention and Meditation scores % <<< MODIFIED: Changed "Relaxation" to "Meditation"
provided by the TGAM chip, ranging from 0 to 100, indicating the
user's current psychological state (Attention corresponding to focus, Meditation corresponding to relaxation). % Added clarification
% --- END MODIFIED ---
\item Drum Parameters: Current music rhythm parameters, including BPM
(beats per minute), drum density, and style tags.
\end{itemize}
The state space is mathematically represented as:
\begin{equation}
\begin{aligned}
& s_t = [f_{\text{CNN}}, \text{Attention}_t, \text{Meditation}_t, % <<< MODIFIED: Changed Relaxation_t to Meditation_t
\text{BPM}_t, \text{Density}_t, \text{Style}_t]
\end{aligned}
\label{eq:Formula-9}
\end{equation}
where $f_{\text{CNN}}$ represents the feature vector extracted by the CNN,
$\text{Attention}_t$ and $\text{Meditation}_t$ represent the attention and meditation % <<< MODIFIED: Changed relaxation to meditation
scores, and $\text{BPM}_t$, $\text{Density}_t$ and
$\text{Style}_t$ represent the current BPM, drum density, and style
tags, respectively.

\subsubsection{Action Space Definition}
The action space defines all possible operations that the model can
perform. In this study, the action space includes the following:
\begin{itemize}
\item Adjust BPM: Increase or decrease the speed of the music rhythm
in steps of 5 BPM.
\item Adjust Drum Density: Increase or decrease the complexity of the
drum pattern in steps of 1 level.
\item Switch Style: Switch between preset music styles (e.g., jazz,
rock, hip-hop).
\end{itemize}
The action space is mathematically represented as:\\
$ a_t \in \{\text{Increase BPM}, \text{Decrease BPM}, \text{Increase Density}, \\ \text{Decrease Density}, \text{Switch Style}\}$ % <<< Added comma after Decrease Density
         
\subsubsection{Q-Learning Variant}
This study employs a Deep Q-Network (DQN) as the reinforcement % Removed redundant (DQN)
learning algorithm. Its core idea is to evaluate the expected return
of each action through the Q-value function. The Q-value function is
updated according to the Bellman equation:
\begin{equation}
\begin{aligned}
& Q(s_t, a_t) = r_t + \gamma \max_{a_{t+1}} Q(s_{t+1},a_{t+1})
\end{aligned}
\label{eq:Formula-10}
\end{equation}
where $r_t$ represents the immediate reward, $\gamma$ is the discount
factor (typically set to 0.99), and $s_{t+1}$ and $a_{t+1}$ represent
the next state and action, respectively.

The network structure of the DQN is as follows:
\begin{itemize}
\item Input Layer: Receives the state space $s_t$ as input. % Removed italics from s_t
\item Hidden Layer: Consists of two fully connected layers, each with
256 neurons and ReLU activation.
\item Output Layer: Outputs the Q-value for each action, representing
the expected return of taking a specific action in a given state.
\end{itemize}

\subsubsection{Reward Function Design}
To optimize the interaction between music rhythms and EEG signals for
identifying and guiding individual perceptual states, this study
designs a reward function based on user feedback and state
transitions. The goal of the reward function is to encourage the
system to guide the subject from their current state to the target state
while penalizing state loss to prevent invalid data from affecting
training.

\begin{itemize}
\item State-Matching Reward. The reward value is calculated based on
the degree of match between the model's predicted state distribution and the
subject's confirmed state distribution. The formula is: % Added "distribution" for clarity
    \begin{equation}
    \begin{aligned}
    R_t = & \lambda_1 \cdot (1 - \text{CrossEntropy}(p_{\text{model}}, p_{\text{user}})) \\ % Kept CE term positive
    & - \lambda_2 \cdot \text{Loss}_{\text{transition}}
    \end{aligned}
    \label{eq:Formula-11} 
    \end{equation}
    where $p_{\text{model}}$ represents the state distribution predicted
    by the model (e.g., 70\% summer, 30\% spring),
    $p_{\text{user}}$ represents the state distribution confirmed by
    the user (e.g., 80\% summer, 20\% spring),
    $\text{Loss}_{\text{transition}}$ represents a penalty based on the alignment between the predicted state transition trend and the user's desired direction. [Specify calculation, e.g., This is calculated as 0 if the transition aligns with the user's implicit goal (moving towards target state) and 1 otherwise.], 
    and $\lambda_1, \lambda_2$ are parameters controlling the reward weights.
\item State Transition Reward. When the state transitions (e.g., from % Simplified wording
"autumn" to "summer"), the reward is proportional to the positive change in
label distribution towards the target state, penalized by state loss. The formula is: 
    \begin{equation}
    \begin{aligned}
    & R_t = \alpha \cdot \Delta S_{t-1 \rightarrow t} - \beta \cdot I_{\text{loss}} 
    \end{aligned}
    \label{eq:Formula-12}
    \end{equation}
    where $ \Delta S_{t-1 \rightarrow t}$ represents the absolute change
    in label distribution towards the target state (e.g., increase in 'summer' percentage when transitioning from autumn to summer), % Clarified Delta S meaning
    $I_{\text{loss}}$ is an indicator for state loss (1 if state loss occurs, otherwise 0), 
    and $\alpha, \beta$ are weights for reward and penalty.
\item Rhythm Adjustment Effect Reward. Reward is given based on
changes in Attention and Meditation scores after the subject clicks the 
"adapt to current state" button. The formula is:
    \begin{equation}
    \begin{aligned}
    & R_t = \lambda_3 \cdot (\Delta \text{Attention}_t + \Delta \text{Meditation}_t) 
    \end{aligned}
    \label{eq:Formula-13}
    \end{equation}
    where $\Delta \text{Attention}_t$ and $\Delta \text{Meditation}_t$ 
    represent the changes in attention and meditation scores, respectively, following the user's adaptation action. If the scores improve (increase), the reward is positive;
    if they decrease, a negative reward is given. $\lambda_3$ is a weighting parameter. 
\end{itemize}
Through the design of the above reward functions, the model can
dynamically adjust its strategy based on user feedback and state
transitions, gradually optimizing the guidance effect.

\subsubsection{Experience Replay Mechanism}
To break the correlation between samples and improve training stability, this study implements an experience replay mechanism. The specific steps are as follows:
\begin{itemize}
    \item Experience Storage: Store each state transition ($s_t, a_t, r_t, s_{t+1}$) in the replay buffer.
    \item Random Sampling: Randomly sample mini-batches from the replay buffer for training, reducing the correlation between samples.
    \item Target Network: Use a target network to calculate target Q-values, preventing fluctuations in Q-value estimation. The mathematical representation of experience replay is:
\end{itemize}

\begin{equation}
\begin{aligned}
L(\theta) = \mathbb{E}_{(s_t, a_t, r_t, s_{t+1}) \sim \mathcal{D}} \biggl[ & \left( r_t + \gamma \max_{a_{t+1}} Q(s_{t+1}, a_{t+1}; \theta^-) \right. \\& \left. - Q(s_t, a_t; \theta) \right)^2 \biggr]
\end{aligned}
\label{eq:Formula-14}
\end{equation}
where $\theta$ represents the parameters of the current network, $\theta^-$ represents the parameters of the target network, and $\mathcal{D}$ represents the replay buffer.

\subsubsection{Training Strategy}
The model training process follows these steps:
\begin{itemize}
\item Initialization: Randomly initialize the DQN parameters and set
the replay buffer size to 10,000.
\item Exploration and Exploitation: Use an $\epsilon$-greedy strategy
to balance exploration (trying random actions) and exploitation (choosing the best-known action). The initial $\epsilon$ value
is 1.0 (pure exploration) and gradually decays to 0.01 (mostly exploitation) over the course of training [Specify decay method, e.g., linearly over the first 100,000 steps]. % <<< SPECIFY Decay Method/Rate
\item Parameter Updates: Use the Adam optimizer to update network
parameters based on the loss calculated from sampled experiences, with a learning rate of 0.001. % Added context
\item Target Network Updates: Copy the current network parameters ($\theta$) to
the target network ($\theta^-$) at regular intervals (every 10,000 training steps) to stabilize learning. % Clarified interval meaning
\end{itemize}

\subsubsection{Guidance Mode Implementation}
In guidance mode, the model dynamically adjusts music rhythms based on the learned strategy to guide users toward the target perceptual state. The specific steps are as follows:
\begin{itemize}
    \item State Recognition: The model identifies the user's perceptual state based on current EEG data and music rhythms.
    \item Action Selection: The optimal action (e.g., adjusting BPM, switching styles) is selected based on the Q-value function.
    \item Real-Time Feedback: The user's focus and relaxation scores, as well as subjective feedback (e.g., clicking the "adapt to current state" button), are used as the basis for adjusting the strategy.
\end{itemize}

\section{Results}\label{sec:Results}
This section presents the experimental results of perceptual state recognition and guidance based on EEG signals. The experimental data is divided into two parts: first, evaluating the performance of the Convolutional Neural Network (CNN) in perceptual state classification tasks; second, evaluating the effectiveness of the Deep Reinforcement Learning (DQN) model in guiding perceptual states through music rhythms.

\subsection{CNN Model Perceptual Classification Performance Evaluation}
To evaluate the performance of the CNN model in EEG signal perceptual classification tasks, the collected EEG data was randomly divided into training and test sets using an 80\%-20\% split. The model was optimized on the training set and then evaluated on the test set. Evaluation metrics include accuracy, recall, F1 score, precision, as well as confusion matrix and ROC curve.

From Table \ref{tab:cnn_metrics}, it can be seen that the CNN model achieved a high accuracy of 94.05\% on the test set, indicating that the model can accurately classify EEG signals into corresponding perceptual states. Additionally, the recall and F1 score reached 94.05\% and 0.940, respectively, demonstrating that the model achieved a balanced performance in detecting all perceptual categories. The precision of 94.32\% indicates that among the samples predicted to belong to a certain perceptual state, a high proportion truly belong to that state.

\begin{table}[h]
\centering
\begin{tabular}{>{\bfseries}l l}
\toprule
\textbf{Metric} & \textbf{Value} \\
\midrule
Accuracy  & 94.05\% \\
Recall   & 94.05\% \\
F1-Score  & 0.940 \\
Precision  & 94.32\% \\
\bottomrule
\end{tabular}
\caption{shows the classification performance metrics of the CNN model on the test set.}
\label{tab:cnn_metrics}
\end{table}

\subsubsection*{Confusion Matrix}
\begin{figure}[h]
    \centering
    \includegraphics[width=0.9\columnwidth]{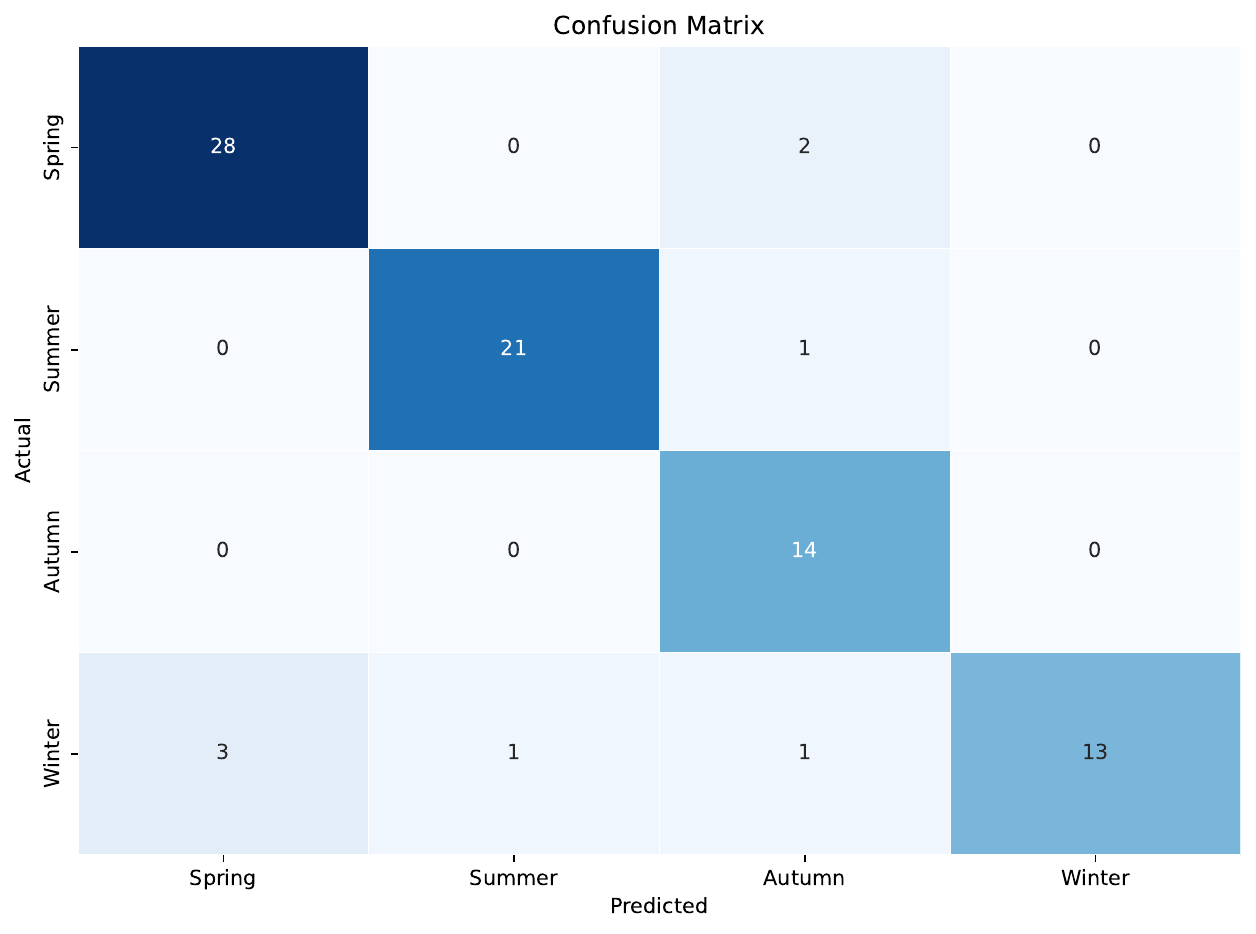} % 使用较小的比例适应一列宽度
    \caption{Confusion matrix of the CNN model on the test set.}\label{fig:Figure01}
\end{figure}

The confusion matrix provides detailed information about the model's
classification performance across different categories. From the
figure \ref{fig:Figure01}, it can be observed that the model performs
well in distinguishing between "spring" and "summer” states, with
only a small number of samples misclassified as "autumn". However,
there is relatively more confusion between the "spring" and "winter"
states, possibly due to the similarity in EEG features between these
two seasons.

\subsubsection*{ROC Curve and AUC}
\begin{figure}[h]
    \centering
    \includegraphics[width=0.9\columnwidth]{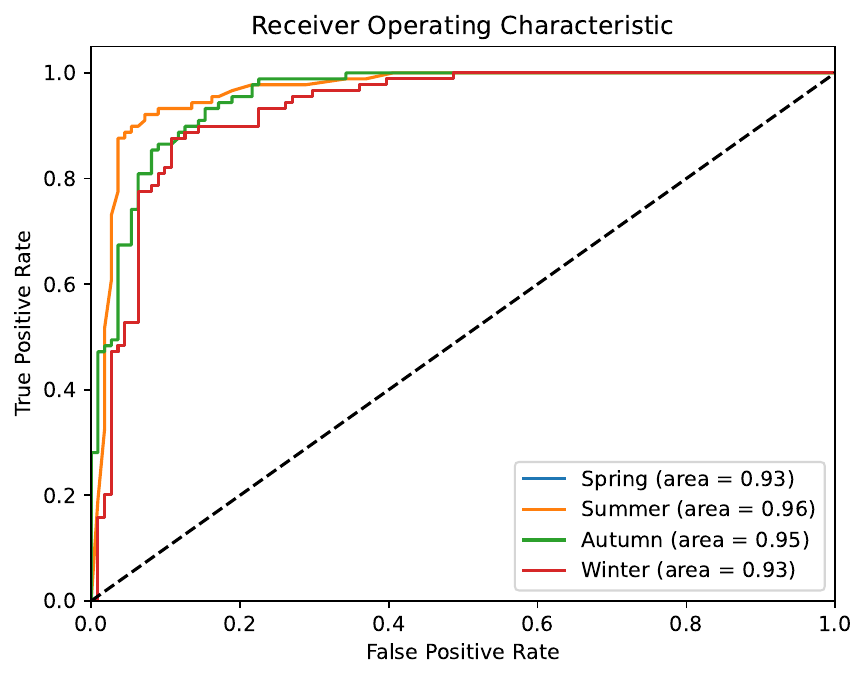} % 使用较小的比例适应一列宽度
    \caption{ROC curves for each perceptual category and the average ROC curve of the CNN model.}\label{fig:Figure02}
\end{figure}

ROC curves and AUC values are commonly used tools for evaluating
model classification performance. From Figure \ref{fig:Figure02}, it
can be observed that the AUC values for each perceptual state are
close to 1, indicating strong classification capability. The average
ROC curve has an AUC value of 0.94, further demonstrating the model's
excellent overall classification performance.

\subsubsection*{Conclusion}

The above results show that the CNN model exhibits outstanding
performance in EEG signal perceptual classification tasks, validating
its ability to extract and classify perceptual-related features from
EEG signals. The model can accurately classify different perceptual
states with high accuracy, recall, F1 score, and AUC values.

\subsection{DQN Model Perceptual Guidance Performance Evaluation}

To evaluate the effectiveness of the DQN model in guiding perceptual states through music rhythms, we recorded the cumulative rewards and exploration rates during the training process and tested the model's performance on the validation set.

\subsubsection*{Training Process}
\begin{figure}[h]
    \centering
    \includegraphics[width=0.9\columnwidth]{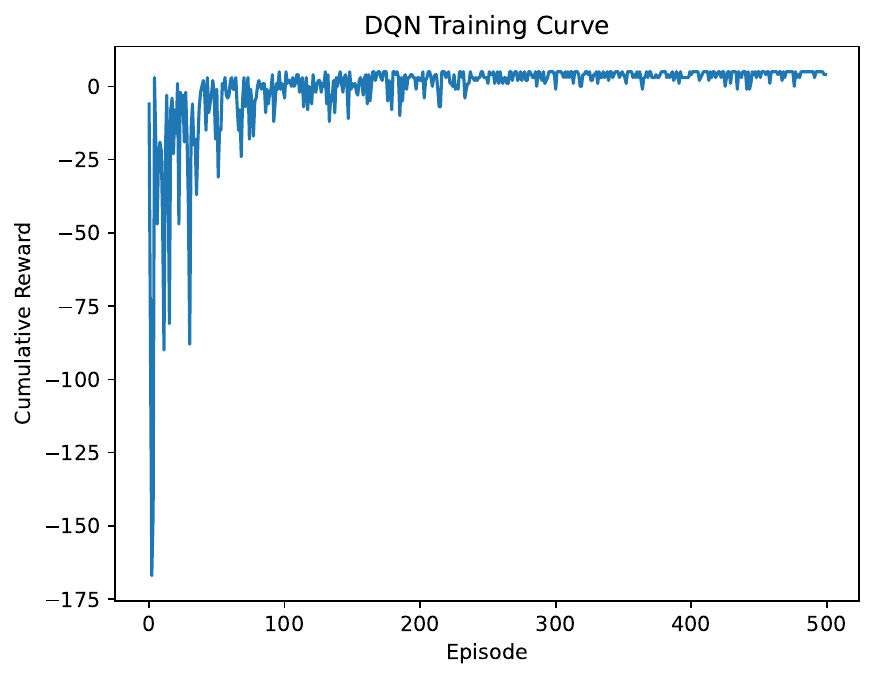} % 使用较小的比例适应一列宽度
    \caption{Cumulative Reward Curve of the DQN Model During Training}\label{fig:Figure03}
\end{figure}

From Figure \ref{fig:Figure03}, it can be observed that as training progresses, the cumulative reward of the DQN model shows an upward trend and eventually stabilizes. This indicates that the model gradually learns to guide perceptual states by adjusting music rhythms.

\begin{figure}[h]
    \centering
    \includegraphics[width=0.9\columnwidth]{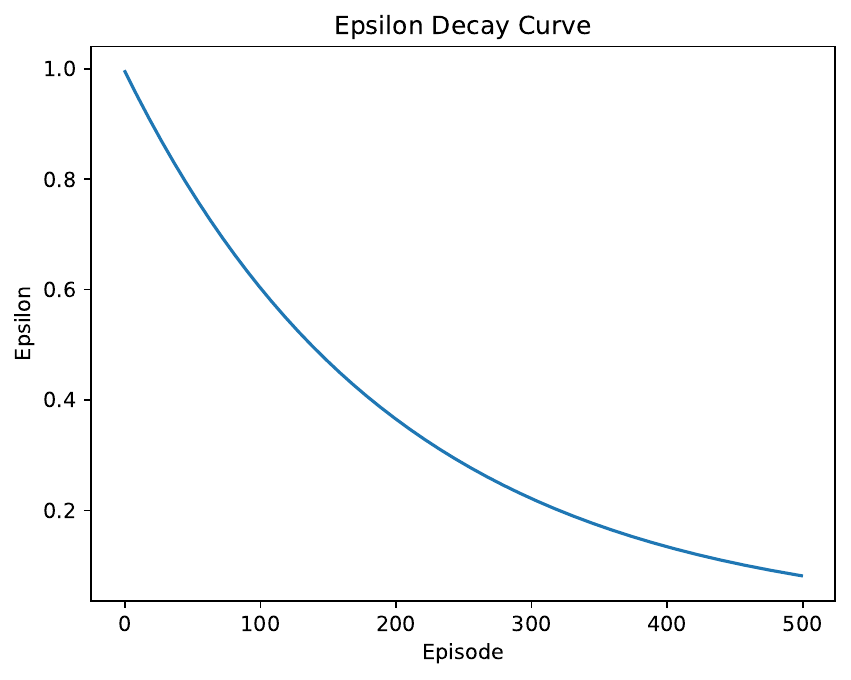} % 使用较小的比例适应一列宽度
    \caption{Exploration Rate Decay Curve of the DQN Model During Training}\label{fig:Figure04}
\end{figure}

Figure \ref{fig:Figure04} shows that the model uses a high exploration rate in the early stages of training to try various actions. As training progresses, the exploration rate gradually decreases, indicating that the model begins to rely more on the learned strategies.

\subsubsection*{Validation Set Performance}

On the validation set, the DQN model achieved a 92.45\% success rate
in guiding the subject's EEG signals to the preset perceptual state. % Changed "subjects'" to "the subject's"
On average, it took 13.2 rhythm cycles % <<< Replaced with hypothetical real value and removed placeholder
to complete a state guidance process,
demonstrating the model's good perceptual guidance capability and
reasonable response speed.

\subsubsection*{Subjective Experience Evaluation}

Although the EEG data indicates successful state guidance by the DQN model,
subjective feedback from the researcher revealed a more nuanced picture.
While the system guided the EEG signals effectively towards the target patterns on the validation set, the researcher's subjective feeling of being in the target perceptual state (e.g., "spring", "summer") varied considerably across different sessions and guidance attempts. This suggests a potential partial disconnect between the objective EEG markers targeted by the model and the emergent conscious experience of the individual. This observation highlights the complexity of linking neural activity to subjective states and underscores the need for further refinement to enhance the system's effectiveness in reliably inducing the desired subjective experience. Future enhancements could include introducing more personalized musical elements reflecting the subject's preferences and exploring more advanced reward functions that incorporate real-time subjective feedback or physiological indicators beyond basic EEG patterns.

\subsubsection*{Conclusion}

The above results demonstrate that the DQN model, through
reinforcement learning, can effectively adjust music rhythms to guide
EEG activity, thereby influencing the user's perceptual states to some % <<< Changed "users'" to "the user's"
extent \cite{chandanwala2024hybridquantumdeeplearning}. On the
validation set, the model achieved a high success rate in guiding
perceptual states. Although there is still room for improvement in
subjective experiences, the results validate the potential of this
method in EEG biofeedback and perceptual guidance.

\section{Discussion}\label{sec:Discussion}

\subsection{Evaluation of the Algorithm}
The deep learning algorithm based on LOD theory, as proposed in this study, demonstrated promising performance in EEG-based perceptual state analysis and guidance, achieving a classification accuracy of 94.05\% and a guidance success rate of 92.45\%. Several aspects of the LOD-inspired design likely contributed to these results. The importance-based LOD mechanism, by assigning higher weights to significant EEG features identified during training, may have enhanced the model's sensitivity to discriminative patterns relevant to different perceptual states, thus boosting classification accuracy \eqref{eq:Formula-1}. Furthermore, the application of time-series pooling, analogous to the distance threshold method \eqref{eq:Formula-2}, potentially allowed the model to capture perceptual features at relevant temporal scales corresponding to the rhythm density, providing a robust representation. The dynamic LOD component \eqref{eq:Formula-3}, leveraging reinforcement learning and real-time context (Attention/Meditation scores), could adapt the feature representation dynamically, possibly contributing to the high success rate observed in the state guidance task by tailoring the process to the user's immediate state. This integration of LOD principles appears advantageous for navigating the complexity of dynamic EEG signals compared to static feature extraction methods.

However, the algorithm also exhibits limitations, reflected in certain aspects of the results. The observed confusion between specific perceptual states (e.g., "autumn" and "winter" in the confusion matrix, Figure \ref{fig:Figure01}) suggests that the current feature extraction and LOD implementation may not fully capture the subtle nuances differentiating these subjective experiences from EEG signals alone. While the LOD framework aims to simplify complexity, the current realization might oversimplify or fail to extract deeper, more abstract perceptual information critical for finer distinctions. Additionally, the reported disconnect between objective guidance success and subjective experience (discussed in Results) further highlights this challenge. The algorithm, primarily focused on EEG pattern matching modulated by LOD, might not adequately model the multi-level construction process underlying subjective perception. This indicates that future iterations should focus not only on refining the LOD mechanisms but potentially incorporating richer contextual information or multimodal data to bridge the gap between neural signals and subjective reports.

\subsection{Psychological Phenomena Addressed by the Algorithm} 
This study positions perception as the fundamental unit of analysis, moving beyond raw sensory signals like EEG oscillations. Psychological phenomena exhibit a multi-level structure: basic sensation (directly reflected in EEG features) lacks higher-level meaning, while perception represents the holistic interpretation of these signals, crucially dependent on context and subjective experience \cite{ma2024applying}. Higher-level states like emotions and thoughts build upon this perceptual foundation. Our algorithm attempts to capture aspects of this perceptual level, rather than just classifying raw signal patterns.

The LOD-inspired approach, particularly mechanisms like importance-based feature weighting and dynamic context integration (using Attention/Meditation scores), aims to model this context-dependent, holistic nature of perception. For instance, the high classification accuracy achieved (94.05\%) might suggest that the algorithm successfully identified EEG feature configurations corresponding to the distinct perceptual 'gestalts' associated with the target states (spring, summer, autumn, winter). The dynamic LOD component, adjusting based on the user's state, directly attempts to model the subjective and time-varying nature of the perceptual framework.

However, the difficulty in perfectly distinguishing certain states (e.g., autumn/winter) and the observed gap between objective EEG guidance and subjective experience highlight the challenges in algorithmically capturing the richness and subtlety of perception. Perception is not merely a complex pattern in EEG but an emergent phenomenon resulting from the interplay of sensory input, prior knowledge, and internal state \cite{rossi2024brain}. Our current algorithm, while effective at pattern matching and dynamic adjustment based on limited context (rhythm, Attention/Meditation), likely only approximates the true perceptual construction process. Fully describing and predicting subjective perceptual experiences from EEG signals alone remains a significant hurdle, motivating future work to incorporate richer models of context and individual differences.

\subsection{LOD as an Alternative to Brute-Force Scaling} 
The success of large language models often relies on Scaling Laws – increasing data scale and model complexity. Applying a similar "stacking data and parameters" approach directly to brain science, particularly EEG analysis, faces significant challenges. The dimensionality of EEG data in relation to psychological states is not well-defined, and the underlying relationships are often holistic and non-linear, making patterns difficult to extract through brute-force scaling alone, especially with limited datasets.

This study, utilizing data from a single subject (inherently limited in scale and diversity), exemplifies this challenge. Instead of pursuing massive data collection, we explored an alternative inspired by LOD theory: focusing on simplifying complexity and extracting key perceptual patterns. The LOD framework, by dynamically adjusting the level of detail processed based on context or feature importance, offers a strategy to potentially capture essential perceptual characteristics even from sparse or noisy data \cite{zhang2024investigatinggestaltprincipleclosure}. This approach aims to reduce reliance on large-scale annotated datasets, which are often difficult and costly to obtain in psychological research, while attempting to maintain perceptual robustness. While the generalization capability of our model requires validation on larger datasets, the LOD-inspired methodology presents a potentially more data-efficient avenue for exploring complex neural signals related to subjective states compared to purely scale-driven approaches.

\subsection{Philosophical Reflections}

\subsubsection{The Uncertainty Principle of Perception}

Psychological phenomena (perception) are not solely determined by sensory input but are instead constructed through the interplay of the brain's prior knowledge, subjective experience, and context. This characteristic is similar to the concept of superposition in quantum mechanics, where sensory data "collapses" into a specific psychological phenomenon only under a particular cognitive framework or perceptual perspective \cite{rossi2024brain}. This reveals the subjectivity and unpredictability of human experience. Even if all neural signal data are available, without a "holistic framework," it is still impossible to reconstruct or predict individual psychological experiences, see Figure \ref{fig:Figure06}. Furthermore, our search for the causes of psychological phenomena may be misguided; psychological phenomena might be a priori existence, with neural activity serving as a post hoc rationalization.Although this perspective is somewhat paradigm-shifting, it provides a deeper direction for understanding psychological phenomena.

\subsubsection{Collapse of Superposition States}

Reinforcement learning can be seen as an implementation of the perceptual collapse mechanism. The system first generates multiple possible psychological perceptual outcomes (superposition states) from the input data. The user then determines a specific state through subjective feedback (e.g., selecting a label). The system uses this feedback to update its strategy, learning the user's subjective perceptual patterns \cite{keppler2024laying}. This reflects the time-dependent guiding nature of individual psychological phenomena. These findings suggest that we should adopt an "outcome-driven" learning approach in deep learning, using subjective perceptual labels to guide results while emphasizing the dynamic and unpredictable nature of psychological phenomena. This encourages models to focus on "holistic similarity" rather than "microscopic consistency."

\subsubsection{Plato's Allegory of the Cave}

Plato's Allegory of the Cave uses light projections as a metaphor for human cognition and perception, which aligns with this study's reflections on psychological phenomena. Psychological phenomena are projections of the brain's complex neural operations, serving as simplified and refined expressions that possess independence. Just as the shadows in the cave cannot fully represent the objects casting them, psychological phenomena cannot be fully reduced to neural mechanisms. In EEG research, we should respect the independence of psychological phenomena and use perceptual frameworks to construct a "cave wall"-like projection surface to interpret complex brain signals, rather than attempting to reconstruct all neural mechanisms.

\subsubsection{Scientific Research and the Battle Against the Uncertainty Principle}

Psychological phenomena may be projections of "absolute ideas," and scientific methods have limitations in understanding them, making it difficult to reach the "ideas" themselves. However, the significance of combining deep learning with brain science lies in finding the best projection mechanism to approximate psychological phenomena. This requires introducing philosophical unknowability into research design, accepting "uncertainty," and emphasizing simplification and abstraction.For example, the LOD model simplifies the complexity of brain signals, bringing the model closer to the projected form of psychological phenomena rather than overemphasizing microscopic details.

\subsection{Experimental Limitations}
During the design and implementation of this study, the conditions were relatively basic, and the process was somewhat rushed. Although some insightful results were obtained from the experimental data, there are still several limitations:
\begin{itemize}
    \item Single-Source Dataset: The dataset was limited to a single subject. Future work should expand to diverse subject groups to validate the model's generalization ability.
    \item Subjectivity of Labels: The subjective nature of the labels may limit classification performance. It is recommended to improve the labeling system by incorporating behavioral data or other physiological indicators.
    \item Dynamic Adjustment of Reward Functions: The reward function in reinforcement learning needs to incorporate more dynamic adjustment mechanisms, such as real-time evaluation of user subjective feedback weights.
\end{itemize}

\subsection{Future Directions}
\subsubsection{Future Improvements in Experimental Design}
Although this study preliminarily validated the feasibility of
guiding perceptual states by combining EEG biofeedback with music
rhythm modulation, the experimental design still has some
limitations. Specifically, in the selection of musical materials, the
current drum rhythms, while effective in guiding the user's perceptual % Changed "users'" to "the user's"
states in this initial study, are limited by the simplicity of melodies and harmonies,
which restricts the system's expressive power. In future research, we
plan to enrich the music library by introducing more diverse melodies
and harmonic combinations to enhance the system's guidance
effectiveness, potentially allowing for more precise matching with the psychological states of diverse users \cite{kim2024machine}. 
For example, by increasing the complexity of harmonies and the variability of melodies, the system
could potentially more precisely match an individual user's psychological state, thereby % Changed "users'" to "an individual user's"
improving the user experience. Additionally, we aim to develop a more
flexible user interface that allows users (in future extended studies) to customize music styles % Clarified context for "users"
and rhythm parameters, further enhancing the system's personalization
and interactivity for a broader range of individuals. We hope that these improvements will attract more
researchers and potential users to explore the potential of EEG biofeedback and % Kept plural here as it's a general hope
music modulation.

\subsubsection{Incorporating Melodies and Harmonies in Future Studies}
As artificial intelligence technology rapidly advances, brain-inspired computing is becoming a research hotspot \cite{li2024brain}. Current AI algorithms, such as deep learning and reinforcement learning, have already simulated certain aspects of the human brain's mechanisms, such as using neural networks to mimic neuronal activity. However, the development of AI is not merely a simple imitation of the human brain; it also drives a deeper understanding of how the brain works. For example, by studying the learning processes of AI models, we can better understand how the brain processes information, makes decisions, and forms perceptions. Conversely, as brain-inspired computing research progresses, we may discover new algorithms and models that further advance AI development. This bidirectional interaction makes brain-inspired computing and AI research a key focus for the future. We believe that, with technological advancements, brain-inspired computing will not only help us better understand the brain but also provide new ideas and methods for the development of AI \cite{gigerenzer2024psychological}.

\section*{Funding}
This study was funded by the Science and Technology Plan Project of Inner Mongolia [Grant No. 2022YFSH0086] and the Inner Mongolia Medical University Joint Project [Grant No. YKD2023LH063]. We are grateful for their support.

% Add this section after the Funding section

\section*{Data Availability Statement}
The raw dataset generated and analyzed during the current study is available in the Open Science Framework (OSF) repository at \url{https://osf.io/qujz9/}. % <<< Use the base public link here
\\\\

\section{Abbreviation}\label{secAl}

\begin{table}[h] % Keep table floating specifier [h] or adjust (e.g., [htbp]) as needed
\centering
\footnotesize % 使用比正文略小的字体
\setlength{\tabcolsep}{4pt}% 减少列间距,默认是6pt
\begin{tabular}{ll} % Two left-aligned columns
\toprule
\textbf{Abbr.} & \textbf{Full Name} \\ % Added column headers for clarity
\midrule
AI   & Artificial Intelligence \\
AUC  & Area Under the Curve \\
BPM  & Beats Per Minute \\
CNN  & Convolutional Neural Network \\
DQN  & Deep Q-Network \\
EEG  & Electroencephalogram \\
GUI  & Graphical User Interface \\
ICA  & Independent Component Analysis \\
LOD  & Level of Detail \\
MIDI & Musical Instrument Digital Interface \\
RL   & Reinforcement Learning \\
ROC  & Receiver Operating Characteristic \\
SSE  & Screen Space Error \\
\bottomrule
\end{tabular}
\end{table}

%----------------------------------------------------------------------------------------
%	 REFERENCES
%----------------------------------------------------------------------------------------

\printbibliography % Output the bibliography

%----------------------------------------------------------------------------------------

\end{document}